\documentclass[a4paper]{panl}
\usepackage{cite}
\usepackage{wrapfig}
\usepackage{graphicx}
\usepackage{amssymb}
\usepackage{amsfonts}
\usepackage{amsmath}
\usepackage{longtable}
\usepackage{rotating}
\usepackage{lscape}
\usepackage{epsfig}
\usepackage{multirow}

\usepackage{lineno}

\originalTeX

\begin{document}

\title{Effect of random environment on kinetic roughening:
Kardar-Parisi-Zhang model with a static noise coupled to the Navier-Stokes equation} 
\maketitle 
\authors{N.V.\,Antonov,$^{a,b}$\footnote{E-mail: n.antonov@spbu.ru} P.I.\,Kakin,$^{a, }$\footnote{E-mail: p.kakin@spbu.ru} M.A.\,Reiter$^{a, }$\footnote{E-mail: mikh.reiter@gmail.com}}
\setcounter{footnote}{0}

\from{$^{a}$\,Department of Physics,  Saint Petersburg State University, Universitetskaya nab.~7/9, 199034~St.~Petersburg,~Russia}

\from{$^{b}$\,N.~N.~Bogoliubov Laboratory of Theoretical Physics, Joint Institute for Nuclear Research, Dubna 141980, Moscow Region, Russia}


\begin{abstract}

Kinetic roughening of a randomly growing surface can be modelled by the Kardar-Parisi-Zhang equation with a time-independent (``spatially quenched'' or ``columnar``) random noise. In this paper, we use the field-theoretic renormalization group approach to investigate how randomly moving medium affects the kinetic roughening. The medium is described by the stochastic differential Navier-Stokes equation for incompressible viscous fluid with an external stirring force. We find that the action functional for the full stochastic problem should be extended to be renormalizable: a new nonlinearity must be introduced. Moreover, in order to correctly couple the scalar and velocity fields, a new dimensionless parameter must be introduced as a factor in the covariant derivative of the scalar field. The resulting action functional involves four coupling constants and a dimensionless ratio of kinematic coefficients. The one-loop calculation (the leading order of the expansion in $\varepsilon=4-d$ with $d$ being the space dimension) shows that the renormalization group equations in the five-dimensional space of those parameters reveal a curve of fixed points that involves an infrared attractive segment for $\varepsilon>0$.
\end{abstract}
\vspace*{6pt} 

\noindent
PACS: 05.10.Cc; 05.70.Fh.

\label{sec:intro}
\section*{Introduction}

Growing profile of fluctuating surface becomes rougher with time; this phenomenon is commonly referred to as ``kinetic roughening'' \cite{spoon}. Stochastic growth occurs in a wide variety of physical systems; its examples include spreading of fires, landscape erosion, evolution of bacterial colonies, etc. \cite{Review}. This justifies interest to kinetic roughening as a whole and to the models developed to describe it.

The Kardar-Parisi-Zhang (KPZ) stochastic differential equation is one of these models \cite{KPZ}; as it is a simple semi-phenomenological model with a term standing for the forces akin to surface tension and a competing nonlinear term for lateral growth, it can be considered as a non-equilibrium analog of the Ising model in equilibrium phase transitions.

The KPZ equation is written for the smoothed (coarse-grained) field $h(x)$ that stands for deviation from average height of a growing surface \cite{KPZ}:
\begin{equation}
    \partial_t h = \kappa_0 \partial_i\partial_i h+\frac{\lambda_0}{2} {\partial_i} h\,{\partial_i} h+\eta.
    \label{KPZeq}
\end{equation}
Here $\eta(x)$ is Gaussian random noise with a certain correlation function; $x=\{t,{\bf x}\}$ are time and space coordinates; $\partial_t=\partial / \partial t$, $\partial_i=\partial / \partial {x_i}$ ($i=1,...,d$, and $d$ is the dimension of space) are the corresponding derivatives; $\kappa_0$ and $\lambda_0$ are the coefficients related to surface tension and lateral growth, respectively. Repeated tensor indices imply summation over them, e.g. $\partial_i \partial_i =\sum_{i=1, \dots, d}\partial_i \partial_i$. 

The random noise $\eta$ in (\ref{KPZeq}) can be chosen in different ways. In this paper, we study the KPZ equation with a time-independent (``columnar'' or ``spatially quenched'') noise:
\begin{equation}
\langle \eta(x)\eta(x') \rangle_{\eta} = \delta^{(d)}({\bf x}-{\bf x'}).
\label{statnoise}
\end{equation}
It should be noted that KPZ equation (\ref{KPZeq}) with the static noise 
(in contrast to a white in-time noise)
does not allow for Galilean symmetry; see \cite{KPZStat}. The mean values of the field $h$ and the noise $\eta$ can be simultaneously ignored; see \cite{KPZ}.

Senisitivity of nearly-critical systems to various types of external disturbances (including deterministic and stochastic flows) is well established; see, e.g.~\cite{Ivanov}. Influence of a moving medium on a growing surface described by Eq.~(\ref{KPZeq}) can be taken into account by the ``minimal replacement'' of the derivative $\partial_{t}$ in Eq. (\ref{KPZeq})  with the Lagrangean  (Galilean covariant) derivative:
\begin{equation}
\partial_{t} h \to    \nabla_t h~=~\partial_t h~+~(v_i \partial_i) h.
\label{replacement}
\end{equation}

The velocity field ${\bf v}=\{v_i\}$ can be modelled by the stochastic differential Navier-Stokes equation for incompressible viscous fluid with an external stirring force:  
\begin{eqnarray}
\partial_{t}v_{i} + ( v_j \partial_j) v_{i} = \nu_0 \partial_j\partial_j v_{i} - \partial_i \wp + f_i.
\label{NS}
\end{eqnarray}
Here $\wp$ is the pressure, $\nu_0$ is the kinematic viscosity coefficient and the incompressibility reduces to the transversality condition $\partial_i v_i = 0$. The Gaussian random force $f_i$ is defined by its pair correlation function \cite{FNS}:
\begin{eqnarray} 
\langle f_i (t, {\bf x}) f_j (t',{\bf x}') \rangle_{f} = 
D_0\,\delta(t-t')\, \int \frac{d{\bf k}}{(2\pi)^d}
P_{ij} ({\bf k})\,  \: \mathrm{exp} \, 
i\{{\bf k}({\bf x} - {\bf x'})\} \,,
\label{whn}
\end{eqnarray}
where $P_{ij}({\bf k}) = \delta_{ij} - k_i k_j / k^2$ is the transverse projector, $D_{0}>0$ 
 and $k\equiv |{\bf k}|$. Due to the presence of a nonvanishing mode at ${\bf k}=0$, the 
force (\ref{whn}) involves macroscopic shaking of the ``tank'' containing the fluid.
 
The stochastic problem (\ref{KPZeq})--(\ref{whn}) is studied on the entire $t$ axis with the retardation condition. Due to to the presence of random noises, initial conditions are irrelevant.

Previous research revealed that when the velocity field is described by the Kazantsev-Kraichnan statistical ensemble for an incompressible fluid, a new nonlinear term must be taken into account alongside the nonlinearity of the original equation (\ref{KPZeq}); see \cite{Universe}. 

In this paper, we apply field-theoretic renormalization group (RG) analysis \cite{Book3} to the full problem (\ref{KPZeq})--(\ref{whn}) to study the effects of medium motion on stochastic growth.

\section*{RG analysis of the problem}

Stochastic problem (\ref{KPZeq})--(\ref{whn}) is equivalent to the field-theoretic model with action functional $S(h,h',{\bf v},{\bf v'})$ (see, e.g., Sec. 5.3 in~\cite{Book3}):
\begin{equation}
\begin{gathered}
    S(h,h',{\bf v},{\bf v'})=\frac{1}{2}h' h'+h' \{ -\partial_t h- \alpha_0 (v_i \partial_i) h+\kappa_0\partial_i \partial_i h+\frac{1}{2}\lambda_0\partial_i h \partial_i h\}+ \\+\frac{1}{2} \sigma_0 h' v_i v_i+v'_i \{ -\partial_t v_i-(v_j \partial_j) v_i+\nu_0\partial_i\partial_i v_i\}+\frac{1}{2}v'_i D_0 v'_i,
    \end{gathered} \label{act}
\end{equation}
where $h'$ and ${\bf v'}$ are auxiliary fields; all required integrations are assumed.

Canonical dimension (and ultraviolet divergences) analysis of the action functional\footnote{For detailed description of these procedures see, e.g. Secs.~1.15, 3.2, and 5.15 in~\cite{Book3}.} shows, firstly, that $d=4$ is a logarithmic dimension. Secondly, the factor $\alpha_0$ must be added to the second term in Eq.~(\ref{replacement}) which explains its appearance in the third term in Eq.~(\ref{act}). Indeed, the counterterm $h'(v_i \partial_i) h$ is needed for renormalization but the product $hh'$ and the field ${\bf v}$ are not renormalized. The only way to reconcile these two facts is to introduce an additional parameter $\alpha_0$.

Finally, the new nonlinearity $\sigma_0 h' v_i v_i$ is also mandated by the ultraviolet divergences analysis (see \cite{Universe} for similar case).

The completely dimensionless ratio $u_0=\kappa_0/\nu_0$ should be considered alongside the four coupling constants (expansion parameters in the ordinary perturbation theory) $g_0=D_0/\nu_0^3$, $w_0=\lambda_0/\kappa_0^2$, $z_0=\sigma_0\kappa_0^2$ and $\alpha_0$. 

We calculated the RG constants in the one-loop approximation (to the leading order in small parameter $\varepsilon=4-d$) and then used them to calculate RG functions (anomalous dimensions and $\beta$ functions) with the same accuracy.  

Fixed points of the RG equations provide information about the possible types of asymptotic (long-time and large-distance, critical) behaviour of the correlation functions. Their coordinates are determined by the zeros of the $\beta$ functions. A fixed point is infrared attractive and determines asymptotic critical behaviour, 
when all the eigenvalues $\lambda_i$ of the matrix $\Omega_{ij}=\partial_i \beta_j$ have positive real parts ${\rm Re} (\lambda_k)~>~0$ (here $\beta_j$ is the full set of $\beta$ functions and $\partial_i$ is the set of derivatives over renormalized coupling constants; see, e.g. Sec. 1.42 in \cite{Book3}).  

By analysing the $\beta$ functions for the action functional (\ref{act}), we found (in the one-loop approximation) a curve of fixed points defined by the following relations: 
\begin{equation}
g^* = 8\varepsilon \slash3, \quad w^{*2} = -2 \varepsilon\slash 3, \quad \alpha^{*2} = {u^*(u^*+1) \slash 6}, \quad z^* = -\alpha^{*2}\slash w^*,
\label{fp}
\end{equation}
where $u^* \neq \{0,-1\}$.  

The curve (\ref{fp}) involves an infrared attractive segment $-0.75<u^*<-0.5$ when $\varepsilon>0$: all the fixed points on that segment are attractive simultaneously.

\section*{Conclusion}

We applied the field-theoretic RG analysis to the Kardar-Parisi-Zhang equation (\ref{KPZeq}) with a time-independent random noise (\ref{statnoise}) coupled to the Navier-Stokes equation (\ref{NS}) with an external stirring force (\ref{whn}). We showed that, to be renormalizable, the action functional for the full stochastic problem Eq.~(\ref{act})  must be modified by addition of a new nonlinearity and a new dimensionless factor in the covariant derivative for the scalar field. These effects are absent for a white in-time noise owing to Galilean symmetry of that problem; see \cite{White} and references therein.

The one-loop calculation of the RG functions revealed a curve (\ref{fp}) of fixed points of the RG equations that includes an infrared attractive segment for $\varepsilon>0$.

In this paper, we did not consider marginal values of two dimensionless parameters $u$ and $\alpha$. A system of $\beta$ functions adjusted for those values may possess new infrared attractive fixed points (see, e.g., \cite{Universe}) so it requires special investigation. This work as well as the calculation of the corresponding critical dimensions are underway.

\section*{Acknowledgments}

The authors would like to express their gratitude to the organizers of the conference "Modern Problems of Condensed Matter Theory 2022" for the opportunity to present their research.

The work was supported by the Foundation for the Advancement of Theoretical Physics and Mathematics ``BASIS'' (P.~I.~K., project number 22-1-3-33-1).

\end{document}